\documentclass{PoS}
\usepackage{lineno}
\usepackage{subcaption}

\title{An improved muon track reconstruction for IceCube}

\ShortTitle{An improved muon track reconstruction for IceCube}

\author{
The IceCube Collaboration$^{\dagger}$\\
{$^{\dagger}$ \itshape \href{http://icecube.wisc.edu/collaboration/authors/icrc19_icecube}{http://icecube.wisc.edu/collaboration/authors/icrc19\_icecube}}\\
E-mail: \email{federica.bradascio@desy.de, thorsten.gluesenkamp@fau.de}
}

\abstract{IceCube is a cubic-kilometer Cherenkov telescope operating at the South Pole. One of its main objectives is to detect astrophysical neutrinos and identify their sources. High-energy muon neutrinos are identified through the secondary muons produced via charge current interactions with the ice. The present best-performing directional reconstruction of the muon track is a maximum likelihood method which uses the arrival time distribution of Cherenkov photons registered by the experiment's photomultipliers. A known systematic shortcoming of the prevailing method is to assume a continuous energy loss along the muon track. This contribution discusses a generalized Ansatz where the expected arrival time distribution is parametrized by a stochastic muon energy loss pattern. This more realistic parametrization of the muon energy loss profile leads to an improvement of about 20\% to the muon angular resolution of IceCube.\\

\vspace{4mm}
{\bfseries Corresponding authors:}
{Federica Bradascio$^{1*}$, Thorsten Gl\"usenkamp$^{2*}$}\\
{$^{1}$\itshape DESY, D-15735 Zeuthen, Germany}\\
{$^{2}$\itshape Erlangen Centre for Astroparticle Physics, Friedrich-Alexander-Universit\"at Erlangen-N\"urnberg, D-91058 Erlangen, Germany}
\phantom{\speaker{F. Bradascio, T. Gl\"usenkamp}}
}

\FullConference{36th International Cosmic Ray Conference -ICRC2019-\\
		July 24th - August 1st, 2019\\
		Madison, WI, U.S.A.}

\begin{document}

\section{Introduction}\label{sec:intro}
IceCube is a cubic-kilometer scale Cherenkov telescope operating at the South Pole, detecting neutrinos of all flavors with energies from tens of GeV to several PeV \cite{ic}. 
It consists of 5160 digital optical modules (DOMs), each containing a 10-inch photomultiplier tube (PMT). The PMTs detect Cherenkov light emitted from charged secondary particles created in neutrino interactions. High-energy neutrinos are thus identified through two main channels: tracks and cascades. Tracks are produced by charged-current interactions of $\nu_{\mu}$ and cascades arise from relatively short particle showers induced by charged-current interactions of $\nu_{e}$ and $\nu_{\tau}$, as well as by neutral-current interactions of all neutrino flavors. 
The search of point sources of neutrinos relies on identifying an accumulation of events from a given direction on top of a large background of atmospheric muons and neutrinos. Muon track events are well-suited for these searches as they can be reconstructed with an angular resolution of $\sim1^{\circ}$ \cite{7yearPS}. 
Reconstruction accuracy depends on geometrical factors (where the track goes inside the instrumented volume) as well as the energy of the event.
Improving the angular resolution for track events therefore directly translates into a better discovery potential of the sources of astrophysical neutrinos. 

The existing muon track reconstruction used in IceCube is based on a maximum likelihood method assuming a minimum ionizing energy loss of an infinite muon track \cite{trackreco, splinempe}. It uses high-dimensional tables which are interpolated with penalized B-splines \cite{spline} to obtain the arrival distribution of Cherenkov photons coming from an infinite muon track. This reconstruction is referred to as \emph{SplineReco} in the following. However, starting above $\sim1$~TeV, muons mostly loose energy stochastically via bremsstrahlung, pair production, and nuclear interactions. The effect of these processes is a production of shower-like depositions on top of the track signature. Light created in such stochastic losses has a different emission spectrum, and it influences the photon arrival time distribution. Therefore, these stochastic energy losses should ideally be included in the track parametrization.

In this paper, a new reconstruction is presented which parametrizes the muon trajectory as a segmented track. This reconstruction is referred to as \emph{SegmentedSplineReco} in the following. Some additional features including better numerical stability, energy regularization, simultaneous vertex and energy optimization, exact gradient/hessian calculation and improved uncertainty estimation are compared to the state-of-the-art muon reconstruction used in IceCube.

\section{The SegmentedSplineReco reconstruction}
\label{sec:ssreco}

\emph{SegmentedSplineReco} is a maximum likelihood reconstruction that uses a cascade-segmented muon hypothesis. These cascades contribute to the probability density function (PDF) of the photon arrival times together with a constant DOM-dependent noise term and an infinite minimum ionizing muon track hypothesis as an option. The number of photons and their time arrival distributions are obtained from high-dimensional splines fitted to Monte Carlo simulations using photon ray-tracing. This superposition of different light emitting sources is different than the standard \emph{SplineReco} method which uses a purely minimum ionizing track hypothesis. In order to reduce the impact of this unrealistic hypothesis, \emph{SplineReco} additionally employs several modifications. These include for example the use of effective photon arrival PDFs from averaged stochastic tracks or the incorporation of other energy reconstructions and likelihood interpolation. These modifications are not necessary in \emph{SegmentedSplineReco} since the stochastic losses are now explicitly modeled. 

The reconstruction \emph{SegmentedSplineReco} performs several steps which  are described in the following:

\begin{enumerate}
    \item The initial hypothesis is twofold: (1) a track and (2) an energy loss pattern parametrized by electromagnetic cascades located along the initial track hypothesis. These first guesses are given by previous reconstructions. 
    Alternatively an energy loss pattern can also be determined directly by \emph{SegmentedSplineReco}, in which case only an initial guess of the track direction is required. 
    \item The total PDF of the photon arrival time $t$ at a DOM position in the detector is given by the weighted sum of $n$ PDFs using each cascade segmented as a source of photon emission:
    \begin{equation}
        \label{eq:pdf}
        p(t) = \sum_{j=0}^{n} w_{j}p_{j}(t),
    \end{equation}
    where $w_{j} = \frac{\lambda_j}{\sum_{k=1}^n \lambda_k}$. The parameter $\lambda_j$ denotes the expected number of photons of source $j$ in the given DOM, where the different sources are the electromagnetic cascades, the constant noise contribution, and additionally a minimum ionizing muon if requested. These expectations values for cascades and the muon are again obtained from high-dimensional spline distributions fitted to simulations.
    \item This PDF is then used to define a likelihood function, which is maximized varying the track parameters ($x$, $y$, $z$, $t$, $\phi_1$, $\phi_2$). 
    Three likelihood functions have been implemented:
    \begin{enumerate}
        \item standard unbinned likelihood $\displaystyle \mathrm{L}=\prod_k \prod_i [p_k(t_i)]^{q_i}$,
        \item extended unbinned likelihood $\displaystyle \mathrm{L}=\prod_k \frac{e^{-\lambda_k} {\lambda_k}^{q_k}}{q_k!} \prod_i [p_k(t_i)]^{q_i}$,
        \item unbinned likelihood for first hit per DOM: $\displaystyle \mathrm{L}=\prod_k [p_{k,1}(t_1)]^{q_1}$.
    \end{enumerate}
    The index $k$ runs over all DOMs while the index $i$ runs over all hits for a given DOM $k$. The charge of a hit $i$ is denoted by $q_i$.
    The PDF $p_{k,1}$ in likelihood (c) is derived from $p_{k}$ as described in \cite{trackreco}.
\end{enumerate}

The new reconstruction includes several improvements with respect to the previous algorithm. This includes support for exact gradient and Hessian calculations from the underlying high-dimensional B-splines and the possibility to fit the energies jointly with the track parameters. The latter option is only feasible now with available gradient information due to the rather high-dimensional ($> 100$D) problem. With the default settings of the reconstruction which fix the energy losses and optimize only the vertex and angular parameters of the track, the gradient-based optimization is typically a factor of two faster due to speed-up convergence.

\section{The angular error estimation}
\label{sec:error}
An important requirement for track reconstructions is to provide a 2D uncertainty estimate of the directional reconstruction for each muon event. Since the direction of the muon track is reconstructed using a maximum likelihood method, a profile log-likelihood scan around the best fit direction in two angle coordinates in the sky can be performed. The error estimate is then obtained by a 2D paraboloid fit on sampled points in this 2D profile log-likelihood space \cite{pb}. This is the prevailing standard method (hereafter \emph{Old method}) used with the \emph{SplineReco} reconstruction and it was applied to \emph{SegmentedSplineReco} as well. However, the running time is typically ten times longer than the reconstruction time and it has a $\sim1\%$ failure rate. Therefore, two new alternative methods have been studied to overcome these limitations.

The first approach (\emph{Method 1}) samples the 6D minimum of track parameters (in $x$, $y$, $z$, $\phi_1$, $\phi_2$, $t$) after optimization with a MCMC (Markov Chain Monte Carlo) sampler. An affine-invariant particle-based sampler \cite{emcee} is used for a few thousand steps. In order to decrease the burn-in phase, the sampler is seeded with the covariance structure at the optimum, which is obtained from a previous Hessian evaluation. These sample points in 6D are then fitted again with a paraboloid shape. As a comparison, the \emph{Old method} \cite{pb} performs the paraboloid fit using profile likelihood evaluations in the two angular dimensions. The paraboloid can be used to define a Gaussian approximation to the optimum of the log-likelihood function. Afterwards all parameters except the two angles are integrated out. If the optimum is non-Gaussian, a discrepancy is observed between the MCMC samples and the paraboloid fit. This discrepancy can be used as indicator of non-Gaussianity and in this case only the MCMC samples could be used if desired. 

In the second approach (\emph{Method 2}) the Hessian is calculated and inverted to obtain the covariance matrix at the minimum. As already mentioned, in most cases the optimum is fairly well described by a Gaussian and hence this method saves a significant amount of computing time. 
In particular, the full covariance matrix with respect to the six track parameters and energy parameters of all individual energy losses can be calculated in this way. For computational reasons, this full covariance matrix calculation has nearly no overhead, but it broadens the final uncertainty contours over the two angular dimensions due to the extra marginalization over the energy dimensions.

\section{Results}
\label{sec:results}
The new reconstruction has been applied to two sets of events with different selection criteria. The first dataset contains events that pass quality cuts (so-called NDir/LDir cuts\footnote{The exact cuts are $\mathrm{LDir} \geq 600$ and $\mathrm{NDir} \geq 8$.}, see \cite{7yearPS} for more information) based on \emph{SplineReco} which to some extent mimics events that are usually found on the final analysis selections used in IceCube. In the following they are referred to as \emph{SplineReco-optimized}. Events with large stochastic losses typically obtain low NDir/LDir values with \emph{SplineReco}. These events subsequently do not pass the cuts and should be mostly absent in this selection.
The second dataset (\emph{Starting events}) is based on a geometrical selection and contains only muon tracks that start in the detector volume and have a minimal track length of $400$ m.

\begin{figure*}
    \centering
    \includegraphics[width=0.8\textwidth,clip]{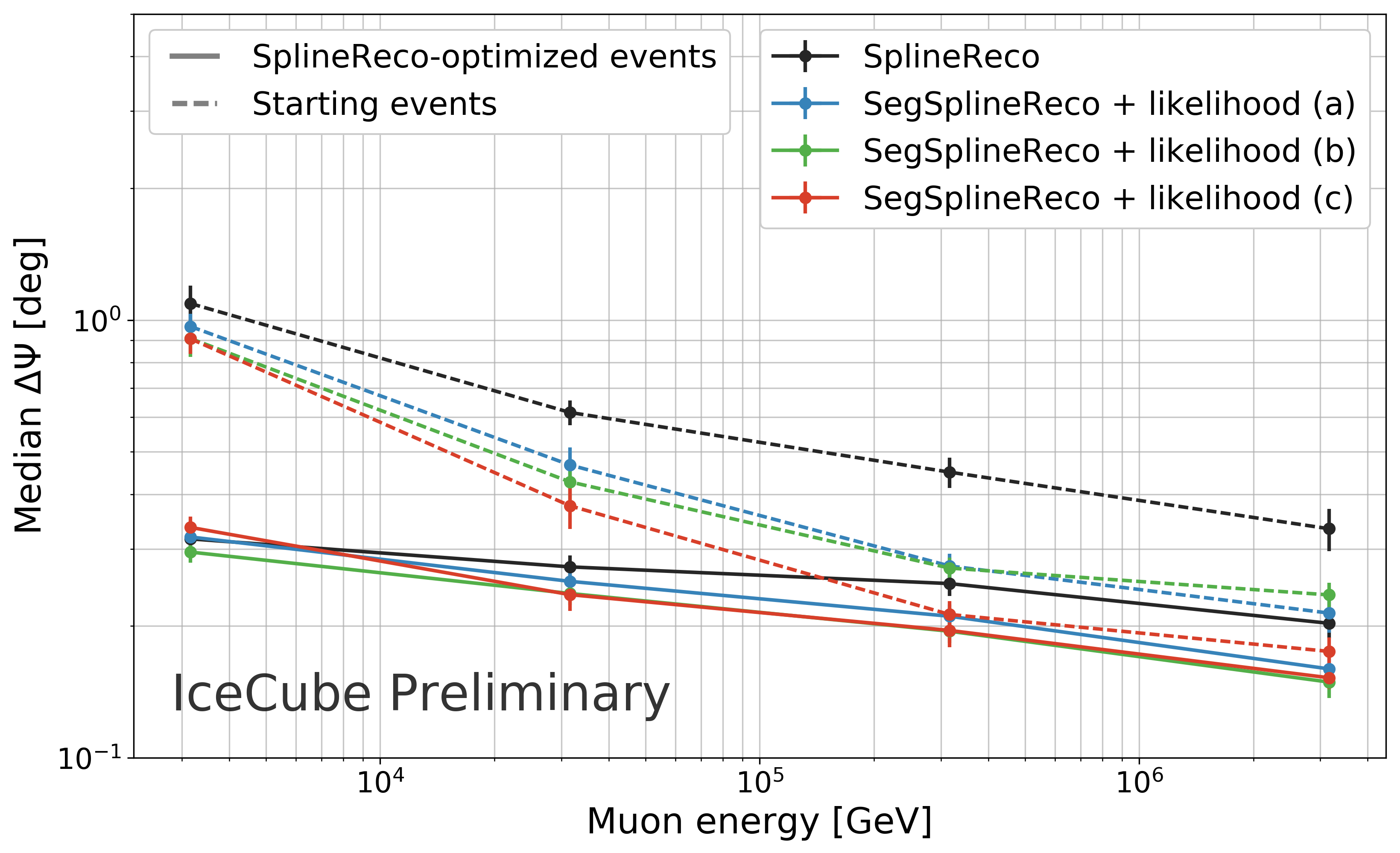}
    \caption{\label{fig:ang_res}Median angular resolution as a function of MC muon energy for two IceCube all-sky simulation: \emph{SplineReco-optimized} events (solid lines) and starting tracks (dashed lines). The \emph{SegmentedSplineReco} reconstruction is compared to \emph{SplineReco} (black line). The three different likelihood models for \emph{SegmentedSplineReco} are compared: (a) the standard unbinned likelihood (blue line), (b) the extended unbinned likelihood (green line) and (c) the unbinned likelihood for first hit per DOM (red line). The statistical error on the median is calculated using bootstrapping.}
\end{figure*}
\begin{figure*}
    \centering
    \includegraphics[width=0.8\textwidth,clip]{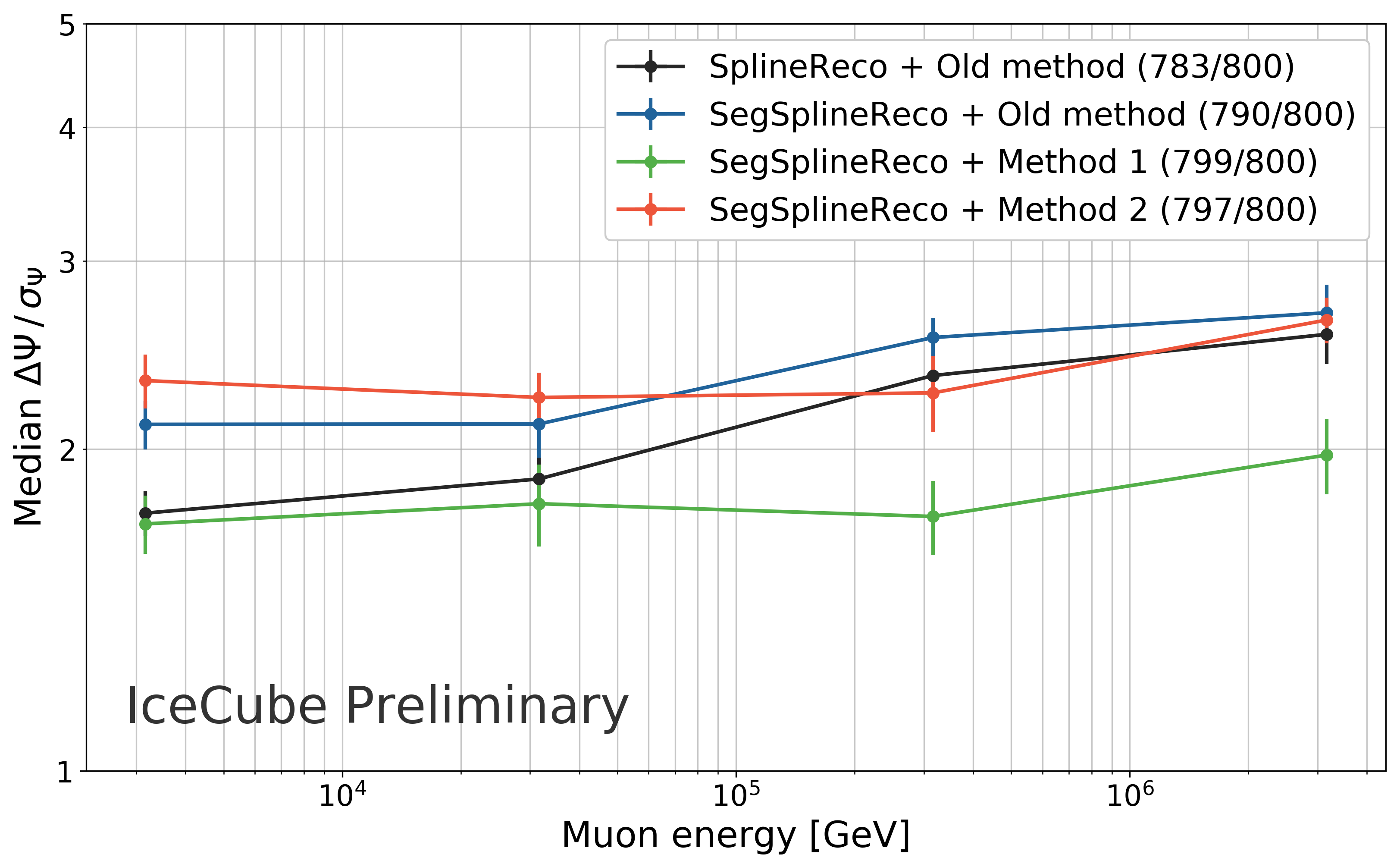}
    \caption{Comparison of different angular error estimators for the \emph{SegmentedSplineReco} reconstruction with likelihood model (c) of the \emph{SplineReco-optimized} events. The median of the pull ($\Delta \Psi / \sigma_{\Psi}$) is shown as a function of the MC muon energy. The statistical error on the median is calculated using bootstrapping. The parenthesis contain the successful reconstruction count.}
    \label{fig:ang_err}
\end{figure*}

Figure \ref{fig:ang_res} shows the median angular error $\Delta\Psi$ of \emph{SegmentedSplineReco} compared to \emph{SplineReco} for both datasets. The input hypothesis for \emph{SegmentedSplineReco} is the reconstructed track from \emph{SplineReco}, and the initial guesses for the stochastic losses are defined in 10 m spacing and fitted once before the actual minimization using the extended unbinned likelihood (b). The track optimization is then performed using fixed cascade energies for the three different likelihood formulations. All these implementations improve the angular resolution with respect to \emph{SplineReco}. Overall, likelihood (c) yields consistently the best results. For the starting events the new reconstruction shows up to factor of 2 improvement in median angular resolution. This is expected because starting events often have a substantial initial hadronic shower loss which is not well captured by \emph{SplineReco}. The result indicates that \emph{SplineReco} as a seed is not optimal for these events, and subsequently explains the large spread for the different likelihoods of the new reconstruction. Likelihood  (c) can mitigate model discrepancies to some extent, and the seed from \emph{SplineReco} is often so far off from the true direction that the initial energy fit results in a strongly biased energy loss model. The results are expected to further improve with iterative energy and subsequent track parameter fitting, in particular for likelihood (a) and (b) that rely stronger on a good model description.

Figure \ref{fig:ang_err} shows the pull ($\Delta\Psi/\sigma_{\Psi}$), defined as the angular difference  divided by the estimated angular difference, using likelihood (c). The standard paraboloid fit and the two new proposed methods were used to determine the angular uncertainty per event. The new methods show flat behavior with respect to the muon energy, except potential small deviations towards the highest energies. The standard paraboloid fit shows stronger energy dependence, in particular for the current \emph{SplineReco} reconstruction. Even with the new methods the uncertainty $\sigma_{\Psi}$ is still understimated. This is expected since a fixed energy loss spacing might not perfectly describe the true loss distribution and the individual energies are fixed. Furthermore, the new methods fail less often. Failure in these cases means the final covariance matrix is not positive definite, indicating the optimizer did not end up at a well-defined local optimum. In these cases it is still possible to resort to the MCMC samples if \emph{Method~1} is being used.

\section{Conclusions and outlook}
An improved high-energy track reconstruction for IceCube, \emph{SegmentedSplineReco}, has been presented. This method models the stochastic energy loss profile more accurately than the prevailing muon reconstruction \emph{SplineReco}.
\emph{SegmentedSplineReco} improves the angular resolution of IceCube by $10\%$ at 100~TeV and $20\%$ at 1~PeV with respect to \emph{SplineReco} on a sample that has been tuned to \emph{SplineReco} quality parameters. A similar improvement is expected when the new reconstruction is applied to existing finished event selections that have been used for muon neutrino point source searches, for example the selection that led to the recent $3.5 \sigma$ neutrino excess from the direction of \emph{TXS 0506+056} \cite{txs}. The new reconstruction has also been applied to geometrically selected starting events, for which an improvement of the median resolution of up to a factor of 2 is observed. In addition to the better angular resolution, the per-event uncertainty estimators have been improved with lower failure rates and better energy-dependent behavior.
Thus, including the new reconstruction as early as possible in the event selection will be interesting for future IceCube data selections, in particular for starting events, which are a key component of astrophysical neutrinos in the Southern Sky.

\end{document}